\documentstyle[12pt]{article}
\begin{document}
\begin{flushright}
hep-th/0309186\\
LTPh, JINR-1991\\
SNBNCBS-2003
\end{flushright}
\vskip 3.5cm
\begin{center}
{\bf \Large { Deformed Traces and Covariant Quantum Algebras\\
for Quantum Groups $GL_{qp} (2)$ and $GL_{qp}(1|1)$ }}

\vskip 3.5cm

{\bf A.P.Isaev and R.P.Malik
\footnote{ E-mail address: malik@boson.bose.res.in  }
\footnote{Permanent Address: S. N. Bose National Centre for Basic Sciences,
Block-JD, Sector-III, Salt Lake, Kolkata, West Bengal, India.}}\\
{\it Laboratory of Theoretical Physics,} \\
{\it JINR, Dubna, P. O. Box 79, Head Post Office, 101000 Moscow, Russia} \\

\vskip 3.5cm

\end{center}

\noindent
{\bf Abstract}:
The $q$-deformed traces and orbits for the two parametric quantum groups
$GL_{qp}(2)$ and $GL_{qp} (1|1)$ are defined. They are subsequently used
in the construction of $q$-orbit invariants for these groups. General
$qp$-(super)oscillator commutation relations are obtained which remain
invariant under the co-actions of groups $GL_{qp}(2)$ and $GL_{qp}(1|1)$.
The $GL_{qp}(2)$-covariant deformed algebra is deduced in terms of the
bilinears of bosonic $qp$-oscillators which turn out to be the central
extension of the Witten-type deformation of $sl(2)$ algebra. In the case
of supergroup $GL_{qp}(1|1)$, the corresponding covariant algebras
contain the $N = 2$ supersymmetric quantum mechanical subalgebras.\\

\baselineskip=16pt

%\vskip 1cm

\newpage

\noindent
Symmetry groups and symmetry algebras have played a very important role
in the development of modern theoretical physics. Quantum groups,
which are deformations (and, in some sense, a generalization) of
usual groups have attracted a great deal of interest since the seminal
papers by Drinfeld [1], Jimbo [2], Faddeev et al. [3], Woronowicz [4]
and Manin [5]. These deformed (super)groups present the examples of
Hopf algebras and have found application in as diverse areas of physics
and mathematics as nonlinear integrable models, statistical mechanics,
conformal field theory, knot theory and solutions of Yang-Baxter equations,
etc., (see, e.g., refs. [6-10] and references therein).

The general quantum deformations of Lie (super)groups or (super)algebras
are the multi-parameter deformations. For instance, the general quantum
deformation of $GL(N)$ has $\frac{1}{2} N (N - 1)$ deformation parameters
[5]. The simple example of the two-parameter quantum deformation of
$GL(2)$ and its differential calculus have been considered in ref. [11].
Following the method of graded tensor product [12], the two-parameter
deformation of the supergroup $GL(1|1)$ has been discussed [13].

Recently the idea of quantum orbits, the one-parameter deformed quantum 
trace, its subsequent application to the construction of $q$-deformed
algebras and the formulation of $q$-deformed Yang-Mills theory have been
developed in ref. [14]. The purpose of our present paper is to define
the quantum trace and quantum orbits for the two parameter groups
$GL_{qp} (2)$ and $GL_{qp}(1|1)$. Further, we obtain the invariants of 
the orbits of these groups and demonstrate that these can be succinctly 
expressed in terms of the deformed traces. Following the approach of
ref. [15], we construct the (super)oscillator algebras which are
covariant under the action of the $GL_{qp}(2)$ and $GL_{qp}(1|1)$
groups and show that the bilinears of these (super)oscillators form
the one-parameter deformed covariant algebra in the case of the
$GL_{qp}(2)$ group and the covariant extensions of the $N = 2$
supersymmetric quantum mechanical algebras for the quantum group 
$GL_{qp}(1|1)$. The more interesting case of $GL_{qp}(2)$ leads to the
construction of the central extension of the Witten-type $q$-algebra
$U_{q}(sl(2))$ [16]. This algebra can be considered as the
``adjoint representation'' of the quantum group $GL_{qp}(2)$. Note that,
in the paper [21], the algebra $U_{pq}(gl(2))$ has been
found  as the dual to the quantum group $GL_{qp}(2)$.

Following Manin's quantum hyperplane approach [5] to the general
construction of quantum groups, it can be shown that the $2 \times 2$
$GL_{qp}(2)$ matrix $T_{ij} =
\left (\begin{array} {cc}
a & b\\ c & d\\
\end{array} \right )$ with the noncommuting elements $a,b,c$ and $d$
exhibits different braiding relations in rows and columns as given by [11]
$$
\begin{array}{lcl}
a  b = p b a, \qquad c d = p d c, \qquad a c = q c a, \qquad
b d = q d b,
\end{array} \eqno(1a)
$$
and other relations are
$$
\begin{array}{lcl}
 b c = {\displaystyle \frac{q}{p}} c b, \qquad 
a d - d a = (p - q^{-1})\; b c = (q - p^{-1})\; c b,
\end{array} \eqno(1b)
$$
where $q, p \in {\cal C}/{\{0\}}$. It is easy to note that the one-parameter 
quantum group $GL_{q} (2)$ corresponds to the special case of (1a)
and (1b) when $q = p$. The inverse quantum matrix $(T_{ij}^{-1})$ is
defined as follows [11]
$$
\begin{array}{lcl}
T_{ij}^{-1} = {\cal D}^{-1}\;
\left (\begin{array} {cc}
d & - q^{-1} b\\ - q c & a\\
\end{array} \right ) \equiv
\left (\begin{array} {cc}
d & - p^{-1} b\\ - p c & a\\
\end{array} \right ) \; {\cal D}^{-1},
\end{array} \eqno(2)
$$
where the quantum determinant ${\cal D} = a d - p b c = d a - q^{-1} b c
= a d - q c b = d a - p^{-1} c b$ is not the central element of the
algebra (1a), (1b) when $p \neq q$, but it obeys the following relations [11]:
$$
\begin{array}{lcl}
&& a \; ({\cal D}, {\cal D}^{-1}) = 
 ({\cal D}, {\cal D}^{-1}) \; a, \qquad 
b \; ({\cal D}, {\cal D}^{-1}) = 
 ({\displaystyle \frac{q}{p}}\; {\cal D}, 
{\displaystyle \frac{p}{q}}\; {\cal D}^{-1}) \; b, \nonumber\\
&& c \; ({\cal D}, {\cal D}^{-1}) = 
 ({\displaystyle \frac{p}{q}}\; {\cal D}, 
{\displaystyle \frac{q}{p}}\; {\cal D}^{-1}) \; c, \qquad
d \; ({\cal D}, {\cal D}^{-1}) = 
 ({\cal D}, {\cal D}^{-1}) \; d.
\end{array} \eqno(3)
$$
Now let us introduce a $2 \times 2$ quantum matrix $E_{ij}$ with 
noncommuting elements. The following transformations of $E_{ij}$
$$
\begin{array}{lcl}
E_{ij} \rightarrow T_{ik}\; E_{kl}\;T_{lj}^{-1},
\end{array} \eqno(4)
$$
define, for all possible $T_{ij} \in GL_{qp}(2)$, the quantum orbit
in the space of $2 \times 2$ $q$-matrices $E_{ij}$ if the elements
of this matrix commutes with that of $T_{ij}$ (i.e. $[T_{ij}, E_{kl} ] = 0$).
Since the elements of $T$ are noncommuting objects, the usual trace of the
matrix $E$ is not invariant under transformations (4). However, it turns
out that the expression (with $r = \surd (qp)$) 
$$
\begin{array}{lcl}
tr_{qp} (E) = tr_{qp} (T E T^{-1}) = r^{-1}\; E_{11} + r\; E_{22},
\end{array} \eqno(5)
$$
remains invariant under (4). We will call this
the quantum $(qp)$-trace. It is straightforward to see that the case
$ q = p$ in (5) yields the one-parameter trace of ref. [14] and $q = p =1$
corresponds to the usual undeformed trace. It may be noticed that
all other invariants for the orbit (4) can be written as
$c^{n} = Tr_{qp} (E^n)$.

Now let us construct the two parametric covariant quantum oscillators for
$GL_{qp}(2)$. With this end in mind, we introduce two sets of $q$-oscillators
$A_i$ and $\tilde A_i ( i = 1, 2)$. In the language of differential geometry
on the quantum hyperplane [17], these operators correspond to coordinates
and derivatives. It is clear that the following relations:
$$
\begin{array}{lcl}
A_1 A_2 - q\; A_2 A_1 = 0, \qquad \tilde A_1 \tilde A_2 - p^{-1} \tilde A_2
\tilde A_1 = 0,
\end{array} \eqno(6)
$$
remain invariant under the $GL_{qp}(2)$ transformations
$$
\begin{array}{lcl}
A_i \rightarrow T_{ij}\;A_j, \qquad \tilde A_i \rightarrow \tilde A_j\;
T_{ji}^{-1},
\end{array} \eqno(7)
$$
where the column matrix $A_i = (A_1, A_{2})^T$ and row matrix
$\tilde A_i = (\tilde A_1, \tilde A_2)$. Consistent with relations
(6), the following oscillator algebra also remains invariant under
transformations (7):
$$
\begin{array}{lcl}
&& A_2  \tilde A_1  - {\displaystyle \frac{(\alpha - \beta)} {p}} \tilde A_1
 A_2 = 0, 
\qquad \;\;\;A_1 \tilde A_2   -  {\displaystyle \frac{(\alpha - \beta)} {q}} 
\tilde A_2 A_1 = 0, 
\nonumber\\
&& A_2 \tilde A_2 - \alpha \;\tilde A_2 A_2 
= 1 + \Bigl ( \alpha - {\displaystyle \frac{(\alpha - \beta)}{r^2}} \Bigr )
\; \tilde A_1 A_1, 
\end{array}\eqno(8a)
$$
if we postulate the validity of the following general bilinear oscillator
relation
$$
\begin{array}{lcl}
A_1 \tilde A_1 - \alpha\; \tilde A_1 A_1 = 1 + \beta\; \tilde A_2 A_2.
\end{array} \eqno(8b)
$$
Here $\alpha$ and $\beta$ are the $c$-number parameters which can be fixed 
by  requiring associativity of the oscillator algebra (6) and (8). In fact,
the oscillator algebra (6) and (8) give us the possibility to reorder a
product of oscillators $A_i$ and $\tilde A_i$ such that all the waved 
operators can be brought to the left side of the products. For example,
let us consider the product $A_1 \tilde A_1 \tilde A_2$. There are two
possible ways to reorder this expression, namely;
$$
\begin{array}{lcl}
\bigl (A_1\; (\tilde A_{1} \;\tilde A_2)\bigr ) = 
{\displaystyle \frac{\alpha - \beta}{r^2}}
\; \Bigl [ \;\tilde A_2 + \alpha\; \tilde A_2 \tilde A_1 A_1 + \beta\;
\tilde A_2 \tilde A_2 A_2 \Bigr ],
\end{array}\eqno(9a) 
$$
$$
\begin{array}{lcl}
\bigl ((A_1\; \tilde A_{1}) \;\tilde A_2) \bigr ) = (1 + \beta)\; \tilde A_2
+ \Bigl ( {\displaystyle \frac{(\alpha - \beta)^2}{r^2}} + \alpha \beta
\Bigr )
\;\tilde A_2 \tilde A_1 A_1 + \alpha \beta\; \tilde A_2 \tilde A_2 A_2.
\end{array}\eqno(9b) 
$$
As can be seen, we obtain two different results on the right hand sides
of (9a) and (9b) which are found to coincide only in two cases
$$
\begin{array}{lcl}
(i)\;\; \alpha = 1 /r^2, \;\;\qquad \beta = 0,
\end{array} \eqno(10a)
$$
$$
\begin{array}{lcl}
(ii)\;\;\alpha = 1 /r^2, \;\;\qquad \beta = 
{\displaystyle \frac{(1 - r^2)}{r^2}}.
\end{array} \eqno(10b)
$$
The case (10a) leads to the following oscillator algebra
$$
\begin{array}{lcl}
&& A_2 \tilde A_1 - q\; \tilde A_1 A_2 = 0, \;\;
 A_1 \tilde A_2 - p\; \tilde A_2 A_1 = 0, \;\;
A_2 \tilde A_2 - pq\; \tilde A_2 A_2 = 1 + (pq - 1)\;\tilde A_1 A_1, 
\nonumber\\
&& A_1 \tilde A_1 - pq\; \tilde A_1 A_1 = 1, \qquad
 A_1 A_2 - q\; A_2 A_1 = 0, \qquad
\tilde A_1 \tilde A_2 - p^{-1}\; \tilde A_2 \tilde A_2 = 0, 
\end{array}\eqno(11) 
$$
while the algebra corresponding to (10b) can be obtained from eqns. (11)
by the replacements $ i = 1 \leftrightarrow i = 2$ and
$ q, p \leftrightarrow q^{-1}, p^{-1}$. It is worth noting that the algebras
of the covariant pair of $q$-oscillators [15] can be obtained from (11) by
the substitution $ q = p, \tilde A_i = A_{i}^\dagger$ and the replacements
corresponding to the solution (10b). Here we stress that the procedure
of obtaining conditions (10) is equivalent to the procedure of deducing
and solving Yang-Baxter equations.

It can now be seen from eqn. (7) that the quantum matrix
$$
\begin{array}{lcl}
E_{ij}\; = \; A_{i} \tilde A_j,
\end{array} \eqno(12)
$$
satisfies the transformation law (4) of the quantum orbit. Furthermore,
the invariance of the trace (5) under transformations (4) leads to the
following $GL_{qp}(2)$ invariant Hamiltonian ($H_{qp}$) in the bilinears
of the covariant oscillators
$$
\begin{array}{lcl}
H_{qp} = r^{-1}\; A_1 \tilde A_1 + r\; A_2 \tilde A_{2}.
\end{array} \eqno(13)
$$
From the $q$-oscillator algebra (6), (8), one can see that the Hamiltonian
(13) is related to the trivially $GL_{qp}(2)$ invariant Hamiltonian
($\tilde H$) by the following equation
$$
\begin{array}{lcl}
\tilde H = {\displaystyle {\sum}_{i = 1}^{2}}\; \tilde A_i A_i
= {\displaystyle \frac{1}{r \alpha + r^{-1} \beta}}\;
[ H_{pq} - (r + r^{-1})],
\end{array} \eqno(14)
$$
where $\alpha$ and $\beta$ are defined in eqns. (10).

Now we would like to define the ``adjoint representation'' of the quantum
group $GL_{qp}(2)$ and establish that the corresponding space of 
representation can be realized as a central extension of the Witten-type
$U_{q}(sl(2))$ algebra. For this purpose, the $q$-matrix $E_{ij}$ can be
rewritten in the following form
$$
\begin{array}{lcl}
E &=& {\displaystyle \frac{1}{r + r^{-1}}} \;\Bigl [\;
\left (\begin{array} {cc}
Tr_{qp} (E) & 0\\ 0 & Tr_{qp} (E)\\
\end{array} \right )\;\; +
\left (\begin{array} {cc}
r E_{0} & (r + r^{-1}) E_{12}\\ (r + r^{-1}) E_{21} & - r^{-1} E_{0}\\
\end{array} \right )\; \Bigr ] \nonumber\\
& =& {\displaystyle \frac{1}{r + r^{-1}}}\;
\bigl [ Tr_{qp} (E) + \tilde E \bigr ],
\end{array}\eqno(15)
$$
where $E_{0} = E_{11} - E_{12}$ and
$$
\begin{array}{lcl}
\tilde E =
\left (\begin{array} {cc}
r E_{0} & (r + r^{-1}) E_{12}\\ (r + r^{-1}) E_{21} & - r^{-1} E_{0}\\
\end{array} \right ).
\end{array} 
$$
From expression (15), it is clear that transformations (4) lead to the
following ``four-dimensional representation'' of $GL_{qp}(2)$
$$
\begin{array}{lcl}
\left (\begin{array} {c} E_{0}^{\prime} \\ E_{12}^{\prime} 
\\ E_{21}^{\prime}\\ H_{qp}^{\prime}
\end{array} \right ) = {\displaystyle \frac{1}{{\cal D}}}\;
\left (\begin{array} {cccc}
{\cal D} + (p + q^{-1}) b c & - (q + p^{-1}) a c
& (p + q^{-1}) d b & 0\\ - p q^{-1} b a & a^2 & - p q^{-2} b^2 & 0\\
q p^{-1} c d & - q^2 p^{-1} c^2 & d^2 & o\\
0 & 0 & 0 & {\cal D}\\
\end{array} \right )\;\;
\left (\begin{array} {c} E_{0} \\ E_{12}
\\ E_{21} \\ H_{qp}
\end{array} \right ).
\end{array}\eqno(16)
$$
This representation is reducible because we have one-dimensional $(H_{qp})$
 and three-dimensional $E_{0}, E_{12}, E_{21})$ invariant subspaces. We can
redefine the operators $E_{+} = E_{12} = A_1 \tilde A_2$ and
$E_{-} = E_{21} = A_2 \tilde A_1$ and show that the oscillator relations
(11) lead to
$$
\begin{array}{lcl}
[ H_{qp}, E_{\pm} ] = [ H_{pq}, E_{0} ] = 0,
\end{array} \eqno(17)
$$
$$
\begin{array}{lcl}
&& [ E_{-}, E_{+} ] = \Bigl ( {\displaystyle \frac{r^2 - 1}{r^2 + 1}}
\Bigr ) \; E_{0}^2 + 
\Bigl ( {\displaystyle \frac{1}{r^2}} +
{\displaystyle \frac{r^2 -1}{r^3 + r}} H_{qp} \Bigr )\; E_{0}, \nonumber\\
&& [ E_{\pm}, E_{\mp} ]_(r^{\mp 1}, r^{\pm 1})
\equiv r^{\mp 1}\; E_{\pm} E_{0} - r^{\pm 1}\; E_{0} E_{\pm} = \pm\;
\Bigl ( {\displaystyle \frac{r^2 - 1}{r^2}}  H_{qp} +
{\displaystyle \frac{r^2 + 1}{r^3 }}  \Bigr )\; E_{\pm}.
\end{array} \eqno(18)
$$
This algebra turns out to be invariant under adjoint rotations (16).

For a given algebra, it is of utmost importance to obtain the Casimir 
operator(s) because the eigen values of such operator(s) designate the
representation of the algebra. The $qp$-deformed quadratic Casimir operator
for the algebra (18) turns out to be related to the invariant
$$
\begin{array}{lcl}
c_{2} = Tr_{qp} (E^2).
\end{array} \eqno(19)
$$
In fact, it can be expressed in terms of of the generators of (18) as given by
$$
\begin{array}{lcl}
c = c_{2} - {\displaystyle \frac{H_{qp}^2}{r + r^{-1}}}
= (r^{-1}\; E_{+} E_{-} + r\;E_{-} E_{+} + 
{\displaystyle \frac{E_{0}^2}{r + r^{-1}}}.
\end{array} \eqno(20)
$$
We can rewrite the commutation relations (18) in a concise form as follows
$$
\begin{array}{lcl}
\tilde E_{ij} \tilde E_{jk} = (r + r^{-1})\; c\; \delta_{ik}
- \kappa\; \tilde E_{ik},
\end{array} \eqno(21)
$$
where $\kappa = [ (r^2 - 1)/r^2 ]\; H_{qp} + (r^2 +1)/r^3$, $c$ is the
Casimir operator (20) and matrix $\tilde E$ is defined in (15). Now
the invariance of algebra (18) under adjoint rotation (16) becomes 
transparent in view of the transformation law $\tilde E \rightarrow
T \tilde E T^{-1}$.

It is worth noting that the algebra (18) depends only on a single
parameter $r = \surd (pq)$ and, therefore, is consistent with Drinfeld's
uniqueness theorem (see, e.g., ref. [11]). The algebras (17) and (18),
in terms of the generators $E_{\pm}, E_{0}$ and $H_{qp}$, give us the
$q$-deformation of $gl(2)$. This covariant algebra is the central
extension of the Witten-type deformation of $sl(2)$ algebra if we consider
$H_{qp}$ as the central element. We see, therefore, that the 
``adjoint-representation'' of the group $GL_{qp} (2)$ leads to
the generalization of the algebra obtained in ref. [16]
(see also ref. [14]).

Now we shall discuss the deformed trace and the orbits for the supersymmetric
case of $GL_{qp}(1|1)$. It is known that the $2\times 2$ quantum matrix
$ T_{ij}^{s} =
\left (\begin{array} {cc}
a & \beta\\ \gamma & d\\
\end{array} \right )$
describes the the $GL_{qp}(1|1)$ group, if the noncommuting odd elements
$\beta$ and $\gamma$ and the even elements $a$ and $d$ satisfy different
braiding relations in rows and columns as given by [13]
$$
\begin{array}{lcl}
&& a \beta = p \beta a,\; \qquad \;a \gamma = q \gamma a,\; 
\qquad \;\;\beta \gamma= - \;(q/p)\; \gamma \beta,\; \qquad\;\;\;
d \gamma = q \gamma d, \nonumber\\
&& d \beta = p \beta d,\quad \beta^2 = \gamma^2 = 0, \quad a d - d a
= - (p - q^{-1}) \beta \gamma = (q - p^{-1}) \gamma \beta,\nonumber\\
&& p, q \in {\cal C}/ \{ 0 \}.
\end{array}\eqno(22)
$$
These relations reduce to the one-parameter case of $GL_{q}(1|1)$ in the
limit $p = q$. The antipode ($ S_{ij} = (T^{s})_{ij}^{-1}$ and the quantum
super determinant ${\cal D}^{s}$ ($q$-berezinian) are obtained by applying the
Borel-Gauss decomposition on the matrix $T^{s}$. These are given as 
follows [13,18]
$$
\begin{array}{lcl}
&&(T^{s})_{ij}^{-1} =
\left (\begin{array} {cc}
a^{-1} (1 + \beta d^{-1} \gamma a^{-1}) & - a^{-1} \beta d^{-1}\\ 
- d^{-1} \gamma a^{-1} & d^{-1} (1 - \beta a^{-1} \gamma d^{-1})\\
\end{array} \right), \nonumber\\
&& {\cal D}^{s} = a d^{-1} - \beta d^{-1} \gamma d^{-1}
= d^{-1} a - d^{-1} \beta d^{-1} \gamma.
\end{array}\eqno(23)
$$
Here ${\cal D}^{s}$ is the center for the algebra (22). The quantum orbits for
 for the supergroup $GL_{qp}(1|1)$ can be defined through the transformations
 (4) for a $2 \times 2$ super $q$-matrix $E_{ij}$ whose elements 
(anti-)commute with that of $T_{ij}^{s}$. Even though the elements of
$T_{ij}^{s}$ follow the graded commutation relations (22), the supertrace
$$
\begin{array}{lcl}
Str_{qp} (E) = E_{11} - E_{22} = Str_{qp} [T^{s} E (T^s)^{-1}],
\end{array} \eqno(24)
$$
remains invariant under  transformations (4). It is interesting to note that
eqn. (24) coincides with the supertraces for the undeformed and the
one-parameter deformed supergroup $GL(1|1)$. It can be shown that all
invariants of the quantum ``superorbit'' $(E \rightarrow T^s E (T^s)^{-1})$
can be expressed as
$$
\begin{array}{lcl}
c_{n} = Str_{qp} \;(E^n) = Str\; (E^n).
\end{array} \eqno(24a)
$$
We now introduce the the set of bosonic $(A, \tilde A)$ and fermionic
$(B, \tilde B)$ variables to study the system of $qp$-superoscillators
that are covariant under the co-action of the quantum supergroup
$GL_{qp} (1|1)$. It is straightforward to demonstrate that the relations
$$
\begin{array}{lcl}
 A B = q\; B A, \qquad \tilde B \tilde A = p \;\tilde A \tilde B,
\qquad B^2 = \tilde B^2 = 0,
\end{array} \eqno(25)
$$
remain invariant under the $GL_{qp}(1|1)$ transformations
$$
\begin{array}{lcl}
\left (\begin{array} {c} A \\ B \\
\end{array} \right )
\;\;\rightarrow \;\;
\left (\begin{array} {cc}
a & \beta\\ \gamma & d\\
\end{array} \right )\;\;
\left (\begin{array} {c}
A \\ B \\
\end{array} \right ) \equiv (\;T^s\;) 
\left (\begin{array} {c}
A \\ B \\
\end{array} \right ),  
\end{array}\eqno(26a)
$$
$$
\begin{array}{lcl}
\left (\begin{array} {cc} \tilde A & \tilde B \\
\end{array} \right )
\;&\rightarrow& \;
\left (\begin{array} {cc} \tilde A & \tilde B \\
\end{array} \right )
\left (\begin{array} {cc}
a^{-1} (1 + \beta d^{-1} \gamma a^{-1}) & - a^{-1} \beta d^{-1}\\ 
- d^{-1} \gamma a^{-1} & d^{-1} (1 - \beta a^{-1} \gamma d^{-1})\\
\end{array} \right), \nonumber\\
&\equiv& 
\left (\begin{array} {cc} \tilde A & \tilde B \\
\end{array} \right )\; (\; T^s\;)^{-1}.
\end{array}\eqno(26b)
$$
Consistent withe the oscillator algebra (25), the other general
$qp$-superoscillator relations are as follows
$$
\begin{array}{lcl}
&& A \tilde B  = {\displaystyle \frac{(\lambda + \nu)} {q}} \tilde B A, 
\;\qquad B \tilde A   = {\displaystyle \frac{(\lambda + \nu)} {p}} \tilde A B, 
\nonumber\\
&& A \tilde A - \Bigl ({\displaystyle \lambda + \frac{(\lambda + \nu)}{pq}} 
\Bigr ) \;\tilde A A 
= 1 - \Bigl ({\displaystyle \nu - \frac{(\lambda + \nu)}{pq}}\;\Bigr )
\; \tilde B B, 
\end{array}\eqno(27)
$$
if we postulate the validity of the deformed anticommutator
$$
\begin{array}{lcl}
 \{ B, \tilde B \}_{(1, \nu)} \equiv
B \tilde B  + \nu \tilde B B 
= 1 + \lambda \tilde A A, 
 \end{array}\eqno(28)
$$
where $\lambda$ and $\nu$ are $c$-number parameters which can be determined
by requiring associativity of the algebra (25), (27) and (28). Indeed, we can 
reorder the tilded oscillators to the left of the product 
$A, \tilde A, \tilde B$ in two different ways which lead to two different
results on the right-hand side. To obtain the unique result on the
right-hand side, we have to take
$$
\begin{array}{lcl}
(i) \;\;\;\;\nu = 1, \qquad \lambda = r^2 - 1, \;\qquad
(ii)\;\;\;\; \nu = 1, \qquad \lambda = 0.
\end{array} \eqno(29a,b)
$$
The two parametric quantum superoscillator algebras corresponding to
the solution (29a) and consistent with eqns. (25) are as follows
$$
\begin{array}{lcl}
&&B \tilde A = q \tilde A B, \qquad A \tilde B = p \tilde B A, \qquad
A \tilde A - pq\; \tilde A A = 1, \nonumber\\
&& B \tilde B\; + \;\tilde B B
= 1\; + \;(pq - 1)\; \tilde A A.
\end{array} \eqno(30a)
$$
Similarly, the solution (29b) yields the following superoscillator
relations consistent with eqns. (25)
$$
\begin{array}{lcl}
&&B \tilde A = p^{-1}\; \tilde A B, \qquad A \tilde B = q^{-1} 
\tilde B A, \qquad
B \tilde B + \tilde B B = 1, \nonumber\\
&& A \tilde A\; - \;{\displaystyle \frac{1}{pq}}\; \tilde A A
= 1\; + \;{\displaystyle \frac{(1 - pq)}{pq}} \;\tilde B B.
\end{array} \eqno(30b)
$$
The case $ q = p, \tilde A = A^\dagger, \tilde B = B^\dagger$ for algebras
(30a) and (30b) gives us the known algebras of the covariant pair of
$q$-oscillators of ref. [15]. We stress here again that the procedure of
obtaining conditions (29) is equivalent to the derivation and solution
of the graded Yang-Baxter equations.

We can now see that the super $q$-matrix
$$
\begin{array}{lcl}
E_{ij} \; = \;
\left (\begin{array} {cc}
A \tilde A & A \tilde B\\ B \tilde A & B \tilde B\\
\end{array} \right ),
\end{array}\eqno(31)
$$
satisfies the transformation laws (4) if the covariant superoscillators
obey the $GL_{qp}(1|1)$ transformation laws (26). The invariance of
$qp$-supertrace (24) leads to the emergence of the following bilinear
representation of the invariant Hamiltonian ($H_{qp}^s$) in terms of 
the the super $qp$-oscillators
$$
\begin{array}{lcl}
H_{qp}^s  = A \tilde A - B \tilde B
= {\displaystyle \frac{\nu + \lambda}{qp}}\;
(\tilde A A + \tilde B B).
\end{array} \eqno(32)
$$
The right-hand side of eqn. (32) is trivially supercovariant in view
of the transformations (26).

It is now obvious that the super transformations (4) lead to the
following four-dimensional ``adjoint representation'' of $GL_{qp} (1|1)$
$$
\begin{array}{lcl}
\left (\begin{array} {c} Y^{\prime} \\ E_{12}^{\prime} 
\\ E_{21}^{\prime}\\ {H_{qp}^s}^{\prime}
\end{array} \right ) = \;
\left (\begin{array} {cccc}
1 &  \gamma d^{-1}
& \beta a^{-1} & \beta d^{-1} \gamma a^{-1}\\ 
0 & {\cal D}^s & 0 & - \beta d^{-1}\\
0 & 0 & ({\cal D}^s)^{-1} & \gamma a^{-1}\\
0 & 0 & 0 & 1\\
\end{array} \right )\;\;
\left (\begin{array} {c} Y \\ E_{12}
\\ E_{21} \\ H_{qp}^s
\end{array} \right ).
\end{array}\eqno(33)
$$
Here $Y = (E_{11} + \mu\; E_{22})/ (1 + \mu)$, where $\mu \neq -1$ is
a $c$-number. The representation (33) is reducible because we have a 
one-dimensional invariant subspace with coordinate $H_{qp}^s 
= E_{11} - E_{22}$. To obtain the covariant algebra for the coordinates
of this representation the key ingredient is to represent these 
coordinates in terms of covariant oscillators using the defining
equation (31). Considering the operators
$$
\begin{array}{lcl}
Y = {\displaystyle \frac{A \tilde A + \mu B \tilde B}{1 + \mu}}, \quad
H_{qp}^s  = A \tilde A - B \tilde B, \quad 
Q = E_{12} = A \tilde B, \quad \bar Q = E_{21} = B \tilde A,
\end{array} \eqno(34)
$$
the covariant superalgebra for the case of (30a) is written as
$$
\begin{array}{lcl}
&& [H_{qp}^s , Q] = 0, \qquad [H_{qp}^s, \bar Q] = 0, 
\qquad [H_{qp}^s, Y] = 0, \qquad Q^2 = \frac{1}{2}\; \{ Q, Q\} = 0, 
\nonumber\\
&& \{Q, \bar Q\} = \bigl [\;1 + (r^2 -1)\; H_{qp}^s \;\bigr ]\; H_{qp}^s
= {\cal H}, 
\qquad \bar Q^2 = 
\frac{1}{2}\; \{ \bar Q, \bar Q \} = 0, \nonumber\\
&& [Q, Y] = + \;\bigl [\; 1 + (r^2 -1)\; H_{qp}^s \;\bigr ]\; Q, \qquad
 [Q, \bar Q] = -\; \bigl [\; 1 + (r^2 -1)\; H_{qp}^s\; \bigr ]\; \bar Q.
\end{array} \eqno(35a)
$$
while for the case (30b), we have the following covariant superalgebra
$$
\begin{array}{lcl}
&& [H_{qp}^s, Q] = [H_{qp}^s, \bar Q] = [H_{qp}^s, Y] = 0, 
\qquad Q^2 = \bar Q^2 = 0, \nonumber\\
&& \{Q, \bar Q\} = H_{qp}^s, \qquad [Q, Y] = + Q, \qquad [Q, Y] = - \bar Q.
\end{array} \eqno(35b).
$$
The Casimir operator $c^s$ for the covariant algebras (35a), (35b)
is related to the invariant $c_2$ (24a) and can be expressed as
$$
\begin{array}{lcl}
c^s = c_2 - {\displaystyle \frac{\mu - 1}{\mu + 1}}\;
(H_{qp}^s)^2 = Q \bar Q - \bar Q Q + 2 Y H_{qp}^s.
\end{array} 
$$

It is interesting to note that in the case of the super quantum group
$GL_{qp}(1|1)$, we obtain $N = 2$ supersymmetric quantum mechanical algebras
(with super charges $Q, \bar Q$, and supersymmetric Hamiltonian ${\cal H}$
(35a) for the case (29a) and $H_{qp}^s$ (32) for the case (29b)) as subalgebras
of the covariant algebras (35a), (35b). The generator $Y$ gives the 
extensions of these quantum mechanical superalgebras. The $q$-superalgebra
(35b) (and superalgebra (35a) after rescaling of the generators 
$Q, \bar Q, Y$ is isomorphic to the undeformed Lie superalgebra
$gl(1|1)$. This is in agreement with the one-parameter case [18].

The emergence of the algebras with generators 
$\{Q, \bar Q, {\cal H} (H_{qp}^s)\}$ from the $GL_{qp}(1|1)$ covariant
superalgebras (35a), (35b) tells about the ``hidden $q$-symmetry'' in
supersymmetric quantum mechanical systems with the Hamiltonians
constructed above. It may be a strong physical motivation for the
study of quantum supergroup.

To conclude, it worthwhile to note that the  (super)traces (5) (for
$ p = q$) and (24) can be  obtained as special cases  from the general
supertrace defined for the one-parameter deformed supergroup
$GL_{q} (N|M)$. To obtain such a general supertrace the essential 
ingredients are the relations between deformed (super)traces and
invariant Hamiltonians that are deduced in eqns. (14) and (32). Moreover,
it is also essential to take into account the results of ref. [15],
where one-parameter $q$-oscillator algebras and corresponding
invariant Hamiltonians, in terms of the bilinears, have been obtained
explicitly. Such a general supertrace for the quantum group $GL_{q}(N|M)$
is as follows
$$
\begin{array}{lcl}
Str_{q} (E) &=& Str_{q} (T E T^{-1})
= (q^{(M-N)/2})\;{\displaystyle \sum}_{i = 1}^{N} (q^{-(N + 1) +2i} E_{ii})
\nonumber\\
&-& (q^{(N-M)/2})\; {\displaystyle \sum}_{s = N + 1}^{N + M}\;
(q^{(M + 1) - 2 (s - N)} E_{ss}),
\end{array} \eqno(36)
$$
where $E_{ij}$ is an $(N + M) \times (N + M)$ $q$-supermatrix and the
elements of the quantum matrix $T_{ij} (i, j = 1, 2....N + M)$ generate
the quantum group $GL_{q} (N|M)$. The matrix $T_{ij}$ acts on the coordinates
$(A_{1}, A_{2}.......A_{N}, B_{1}, B_{2}, ......B_{M})$ defined on the 
quantum hyperplane. The basic relations for these coordinates are as follows
$$
\begin{array}{lcl}
A_i A_j = q\; A_j A_i, \; (i < j), \qquad
A_i B_{s} = q\; B_s A_i, \qquad B_r B_s = - q B_s B_r, \; (r < s).
\end{array} \eqno(37)
$$
We would like to emphasize at this juncture that the substitutions
$ N = 2, M = 0$ and $N = 1, M = 1$ lead to the emergence of the expressions
for the (super)traces obtained in (5) (for $p = q$) and (24). Furthermore,
$M = 0$ in eqn. (36) yields the expression for for the $q$-trace presented
in ref. [14] for the quantum group $GL_{q}(N)$. The special case of
this $q$-trace $(N = 3, M = 0$) was used in ref. [19] in the context
[3, 20] of the construction of the central elements for the quantum algebras.

It is possible to extend the covariant  (super)algebras (18) and (35) which
correspond to ``adjoint representation'' (16), (33) and construct the higher
dimensional ``representations'' for the quantum groups $GL_{qp}(2)$,
$GL_{qp}(1|1)$. For this purpose, one has to consider  the following tensors
 which are elements of the enveloping algebra of $qp$-oscillators
$$
\begin{array}{lcl}
E^{i_1, i_2,......i_n; r_1, r_2....r_{m}}_{j_1, j_2,....j_{n};
s_{1}, s_2....s_{m}} = \tilde A_{i_1}......\tilde A_{i_n} B_{r_1}.....B_{r_m}
A_{j_1}.......A_{j_m} B_{s_1}.......B_{s_{m}}.
\end{array} \eqno(38)
$$
The transformations presented in (7) and (26) define the action of
$GL_{q}(1|1)$ and $GL_{q}(2)$ on such tensors and give rise to the higher
dimensional ``representations'' of these quantum groups. The interesting 
non-trivial problem here would be to separate (38) into irreducible 
representations and investigate corresponding covariant infinite-dimensional
algebras. We hope that the notion of the quantum trace will help in the
resolution of this problem.

The authors would like to thank Ch. Devchand and P. Pyatov for fruitful
discussions. One of us (A.P.I.) would like to gratefully acknowledge
valuable discussions with R. Kashaev, J. Lukierski, Z. Popowicz and
V. Tolstoy. \\

\noindent
{\bf Note added:}\\

\noindent
After we communicated this paper to the arXiv, we were informed by
V. K. Dobrev about his paper [21] where the algebra $U_{qp}(gl(2))$,
as the dual to the quantum group $GL_{qp}(2)$, has been constructed.

\end{document}